\documentstyle[pra,aps,epsf]{revtex}
\newcommand{\be}{\begin{equation}}
\newcommand{\ee}{\end{equation}}
\newcommand{\bea}{\begin{eqnarray}}
\newcommand{\eea}{\end{eqnarray}}
\newcommand{\osi}{$^{16}$O}

\newcommand{\ble}{$^{208}$Pb}

\begin{document}
\title{Correlations and the Cross Section of Exclusive ($e,e'p$) Reactions
for $^{16}$O}

\author{K. Amir-Azimi-Nili, J.M. Udias\footnote{Present address:
Institute des Science Nuclaires, 53 Avenue des Martyrs, F-38026
Grenoble CEDEX (France)}, H. M\"uther}

\address{Institut f\"{u}r Theoretische Physik,\\ Universit\"{a}t T\"{u}bingen,
D-72076 T\"{u}bingen, Germany}

\author{L.D. Skouras}

\address{Institute of Nuclear Physics,
N.C.S.R. Demokritos \\Aghia Pa\-ras\-kevi GR 15310, Greece}

\author{A. Polls}

\address{Departament d'Estructura i Constituents de la Mat\`eria\\
Universitat de Barcelona,
E-08028 Barcelona, Spain}

\maketitle
\date{\today}

\begin{abstract} 
The reduced cross section for exclusive ($e,e'p$) reactions has been
studied in DWIA for the example of the nucleus $^{16}$O using a
spectral function containing effects of correlations. The spectral
function is evaluated directly for the finite nucleus starting from a
realistic nucleon-nucleon interaction within the framework of the
Green's function approach. The emphasis is focused on the correlations
induced by excitation modes at low energies described within a
model-space of shell-model configurations including states up to the
$sdg$ shell.  Cross sections for the $p$-wave quasi-hole transitions
at low missing energies are presented and compared with the most
recent experimental data. In the case of the so-called perpendicular
kinematics the reduced cross section derived in DWIA shows an
enhancement at high missing momenta as compared to the PWIA
result. Furthermore the cross sections for the $s$- and $d$-wave
quasi-hole transitions are presented and compared to available data at
low missing momenta. Also in these cases, which cannot be described in
a model without correlations, a good agreement with the experiment is
obtained.
\end{abstract}

\section{INTRODUCTION}

The quasi-elastic ($e,e'p$) reaction constitutes a very well suited
tool to study the limitations of the simple shell-model or
independent-particle model (IPM) of the nucleus. Indeed, it is now
generally accepted that atomic nuclei are many-body systems in which
correlations beyond the mean-field or Hartree-Fock picture play a
significant role.  It has been argued that the
strong short-range and tensor components of realistic nucleon-nucleon
(NN) interactions should induce short-range correlations into the
nuclear wave function. These correlations should give rise to an
enhancement in the momentum distribution of quasi-hole states at high
momenta as compared to the momentum distribution derived from a
Hartree-Fock or mean field description of the nucleus. Therefore
high-resolution $(e,e'p)$ experiments have been performed to determine
the spectral function of nucleons at high momenta leading to the
ground-state or states with low excitation energies in the daughter
nucleus \cite{mainz,adam}.

Microscopic calculations, which account for the effects of these
short-range correlations, indeed predict components in the momentum
distribution at momenta around 2-3 fm$^{-1}$, which are larger by
orders of magnitude as compared to the predictions of Hartree-Fock or
IPM calculations. These high-momentum components, however, mainly originate
from the spectral function at large missing energies, leading to 
configurations of the (A-1) particle system at energies above the
threshold of particle emission for another nucleon.  The
momentum distribution for nucleons at low missing energies, which can be
explored in exclusive ($e,e'p$) reactions, seems not to be very sensitive
to short-range correlations and may rather
well be approximated by those derived from the IPM
\cite{pieper,co,benhar,artur1,wim,polls}. Similar results are also
obtained in the study of the spectral function for nuclear matter
\cite{ramos,benh,vonder,knehr}.

This analysis was essentially confirmed by experiment. The
experimental data for the reduced cross section in the light nucleus
\osi\ at low missing energies could be described reasonable well
within the prediction of an IPM \cite{mainz}. On the other hand, the
momentum distribution for the heavy nucleus \ble\ showed a small
enhancement of high momemtum components at low missing energies as
compared to a Hartree-Fock or IPM prediction \cite{adam}. This
enhancement is not as large as the one predicted from short-range
correlations for the total momentum distribution including the
contributions from high missing energies. It has been argued that this deviation
from the IPM might be due to long-range correlations, which
corresponds to low-energy excitations of the many-body system
\cite{adam,maw}. 

The fact that this enhancement of the momentum distribution is
observed for a heavy nucleus but not for the light nucleus \osi\
supports the idea that the enhancement at small energies may originate
from long-range correlations. The effects of short-range correlations
should be rather independent of the nuclear system under consideration
as these correlations are not very sensitive to the
global structure of the whole nuclear system. Consequently the effects
of short-range correlations should be rather similar for the nuclei
\osi\ and \ble . Contrary long-range correlations could be sensitive
to the whole nuclear system and exhibit different results for different
nuclei. They are related to the excitation modes at low energy and
therefore results derived from nuclear matter, which shows a
continuous single-particle spectrum, can be quite different from those
in finite nuclei, for which the low-energy excitations are rather
sensitive to the shell-structure. Guided by these considerations we
investigated long-range correlation effects in the spectral function
and other related observables directly for the finite nucleus under
consideration \cite{amir}, as a study of nuclear matter may not be
very reliable.

If one wants to study the effects of correlations by a comparison with
experimental data, one has to employ the correlated wave function in 
a complete description of the ($e,e'p$) reaction. Going beyond the
Plane Wave Impuls Approximation the cross section cannot be factorized
any longer into the elementary electron-proton cross section
and the spectral function, 
exhibiting the effects of correlations. The reduced cross section
calculated in the Distorted Wave Impuls Apoproximation (DWIA) may
contain effects of the final state interactions (FSI) and other effects
which cannot be separated in this approach from the effects of
correlations. Therefore a careful analysis has
to be made of the different components entering in the computation of
the reduced cross section.

 Also we want to emphasize that one should investigate the reduced
cross section leading to final states in the daughter nucleus, which are
absent in the IPM. In the case of \osi\ processes of this kind would be
the knock-out of a nucleon leading to the 5/2 and 1/2 states of positive
parity with low excitation energy in $^{15}$N. In the IPM the $d_{5/2}$
and $1s_{1/2}$ states are not occupied and therefore 
the corresponing momentum distributions are not
dominated by the quasi-hole part of the simple shell-model.

To account for long-range correlations a harmonic oscillator basis is
used in order to determine the energies and the mixing of shell-model
configurations. We consider that the low-energy excitation modes are
adequately described within a model-space which includes all orbitals
up to the $sdg$ shell. A finite basis of oscillator
states, however, is not at all appropriate to describe high-momentum
components in the nuclear wave function, since these high-momentum
components will be dominated by the tail of the oscillator basis
states. Therefore, as we will explain below, we have used a mixed
representation of basis states, which considers a shell-model basis to
describe the excitation modes, but a basis of plane-wave states to
determine the spectral function. The Green's function formalism for
the nuclear many-body theory (for a reference see {\em e.g.}\ the
recent review articles \cite{mahaux,wimrev}) will be applied to
evaluate the spectral function.

In Sec. \ref{sec:spec} we will present the main points of the method
we use to compute the correlated spectral function for \osi .
This part is followed in Sec. \ref{sec:red} by a brief summary of the
formalism used to describe the ($e,e'p$) reaction, where Coulomb
distortion effects of the electron wave function and final state
interaction of the outgoing proton are considered. The results for the
reduced cross section for the different approximations and partial
wave transitions are presented in Sec. \ref{sec:res}. At the end we
give a short review of our main conclusions.

\section{THE SPECTRAL FUNCTION}
\label{sec:spec}

The calculation of the cross section for exclusive ($e,e'p$)
reactions requires the knowledge of the hole spectral function. In the
case of finite system it is convenient to introduce a partial wave
decomposition which yields the spectral function for a nucleon in the
single-particle basis with orbital angular momentum $l$, total angular momentum
$j$, isospin $\tau$ and momentum $k$
\begin{equation}
S_{l j \tau} (k,k';E) \quad = \quad \sum_{n} \enspace 
\langle  \Psi_{0}^{A} | a_{k'lj}^\dagger | \Psi_{n}^{A-1} \rangle
\langle\Psi_{n}^{A-1}|
a_{klj\tau} | \Psi_{0}^{A} \rangle \enspace \delta 
(E-(E_{0}^{A}-E_{n}^{A-1}))\, ,
\label{specf}
\end{equation}
where $a_{klj\tau}$ ($a_{k'lj\tau}^\dagger$) denotes the corresponding
annihilation (creation) operator. The state $|\Psi_{0}^{A}\rangle$
refers to the ground state of the target nucleus, while $|
\Psi_{n}^{A-1}\rangle$ is used to identify the $n$th excited
eigenstate of the hamiltonian with one particle removed from the
target nucleus. Hence the hole spectral function in its diagonal form
$S({\bf k},E)$ gives the probability of removing a particle with
momentum ${\bf k}$ from the target system of $A$ particles leaving the
resulting ($A$--1) system with an energy $E^{A-1}=E_{0}-E$, where
$E_{0}$ is the ground state energy of the target. The spectral
function for the various partial waves, $S_{l j \tau} (k,k',E)$ can be
obtained from the imaginary part of the corresponding single-particle
Green's function $g_{lj} (k_1,k_2,E)$. Note that here and in the
following we have dropped the isospin quantum number $\tau$, as we
ignore the Coulomb interaction between the protons and study a symmetric
nucleus with $N=Z$.

To determine the correlated single-particle Green's function one has to
solve a Dyson equation, which using the partial wave representation
appropriate for finite systems can be written as
\begin{equation}
g_{lj}(k_1,k_2;\omega ) = g^{(HF)}_{lj}(k_1,k_2;\omega )
+ \int dk_3\int dk_4 g^{(HF)}_{lj}(k_1,k_3;\omega ) \Delta\Sigma_{lj}
(k_3,k_4;\omega )  g_{lj}(k_4,k_2;\omega ) \,,
\label{dyso}
\end{equation}
where $g^{(HF)}$ refers to the Hartree-Fock propagator and
$\Delta\Sigma_{lj}$ represents contributions to the irreducible
self-energy, which go beyond the Hartree-Fock approximation for the
nucleon self-energy used to derive $g^{(HF)}$. The hole spectral function
$S_{lj}$ can then be calculated easily from the imaginary
part of the single-particle Green's function by
\begin{equation}
S_{lj}(k,k',\omega) \, = \,
\frac{1}{\pi} {\rm  Imag} \bigl[g_{lj}(k,k';\omega )\bigr] 
\, .
\label{modi}
\end{equation}
Although the evaluation of the Hartree-Fock approximation to the
Green's function $g^{(HF)}_{lj}(k_1,k_2;\omega )$, the definition of
$\Delta\Sigma_{lj}$ and the technique used to solve the Dyson
Eq.(\ref{dyso}) have been already discussed in detail in previous
publication \cite{skouras,amir}, we include a brief summary of the
relevant aspects of this method.

\subsection{Model space and effective hamiltonian}

Long range correlations are taken into account by means of the
Green's function approach within a finite model space. This model
space shall be defined in terms of shell-model configurations
including oscillator single-particle states up to the sdg shell. The
oscillator parameter, $b=1.76$ fm, has been chosen appropriate for the
nucleus \osi. This model space does not allow the description of
short-range correlations.  Nevertheless, we also have to take into
account the effects of short-range correlations by introducing an
effective interaction, {\em i.e.}\ a $G$-matrix appropriate for the model
space. This truncation of the Hilbert space into a model space, the
degrees of which are treated explicitly, and the space outside this
model space, which is taken into account by means of effective
operators, is often referred to as a double partitioned Hilbert space
and has been used before for finite nuclei \cite{bar,sko1} and nuclear
matter \cite{kuo3,kuo2}.

The $G$-matrix is determined as the solution of the
Bethe-Goldstone equation
\begin{equation}
{\cal G} = V + V \frac{Q_{\hbox{mod}}}{ \omega - Q_{\hbox{mod}} T
Q_{\hbox{mod}}} {\cal G}\; ,
\label{gmat}
\end{equation}
where $T$ is identified with the kinetic energy operator, while $V$
stands for the bare two-body interaction. For the latter we have
chosen the Reid soft-core potential\cite{reid}. In this equation
the Pauli operator $Q_{\hbox{mod}}$ is defined in terms of our harmonic
oscillator single-particle states. Thus applying $Q_{\hbox{mod}}$ to
two-particle states $\vert \alpha\beta >$ one obtains
\begin{equation}
 Q_{\hbox{mod}} \vert \alpha\beta > = \left\{ \begin{array}{ll}
0 & \mbox{if $\alpha$ or $\beta$ below Fermi level}\cr
0 & \mbox{if $\alpha$ and $\beta$ in model space} \cr
\vert \alpha\beta > & \mbox{elsewhere} \end{array} \right. 
\label{paul}
\end{equation}
The model space used in the Eq. (\ref{gmat} and \ref{paul}) includes all
states up to the sdg-shell. Note that with this definition of
$Q_{\hbox{mod}}$ we ensure that no doublecounting of correlations occurs
between the short-range correlations taken into account in the
${\cal G}$-matrix and the long-range correlations evaluated by means of
the Green's function method within the model-space.

In the solution of the Bethe-Goldstone Eq.(\ref{gmat}) we have chosen a
constant value of $\omega = -30$ MeV for the starting energy. This value is
a reasonable mean value for the sum of two single-particle energies for
hole states in \osi . Clearly, this choice is an approximation introduced
to simplify the calculations, but one has to note that our results do not
depend significantly on the actual value of $\omega$. The use of a constant
starting energy also implies that we do not try to account for a depletion
of the occupation probability due to scattering into states outside the
model space as it has been done {\em e.g.}\ in \cite{yang,allart}.

The matrix elements for ${\cal G}$ are computed in a plane-wave basis for a
specific finite nucleus and a correspondig model-space by expanding the
techniques for solving the Bethe-Goldstone equation for finite systems as
described in Ref. \cite{sauer}.

\subsection {Nucleon Self-Energy and Green's Function}

The calculation of the self-energy is performed in terms of
two-particle states, characterized by  single-particle momenta in the
laboratory frame. Such a antisymmetrized 2-particle state would be
described by quantum numbers such as 
\begin{equation}
|k_{1}l_{1}j_{1}k_{2}l_{2}j_{2}JT>
\label{labf}
\end{equation}
where $k_{i}$, $l_{i}$ and $j_{i}$ refer to momentum and angular
momenta of particle $i$ whereas $J$ and $T$ define the total angular
momentum and isospin of the two-particle state. The transformation
from the relative and c.m. coordinates, in which the matrix elements
of ${\cal G}$ are defined to the states displayed in Eq.(\ref{labf})
can be made by use of the well known vector bracket transformation
coefficients \cite{wong,bonats}.

Performing an integration over one of the $k_{i}$, one obtains a
2-particle state in a mixed representation of one particle in a bound
harmonic-oscillator state while the other is in a plane-wave state
\begin{equation}
|n_{1}l_{1}j_{1}k_{2}l_{2}j_{2}JT> \quad = \quad
\int\limits_{0}^{\infty} dk_{1} \enspace k_{1}^{2} \enspace
R_{n_{1}l_{1}} (b\, k_{1}) \enspace
|k_{1}l_{1}j_{1}k_{2}l_{2}j_{2}JT>.
\label{mixed}
\end{equation}
Here $R_{n_{1}l_{1}}$ stands for the radial oscillator function.
An oscillator length $b = 1.76$ fm ( $ \hbar\omega = 13.3$ MeV) has been
selected, which is an appropriate value to describe
the bound single-particle states in $^{16}$O. Now
with the help of eqs.(\ref{labf} - \ref{mixed}) we can write down our
Hartree-Fock approximation for the self-energy in momentum
representation 
\begin{equation}
\Sigma^{HF}_{l_1j_1} (k_1,k'_1) =
\frac{1}{2(2j_1+1)} \sum_{n_2 l_2 j_2 J T} (2J+1) (2T+1)
\left\langle k_1 l_1 j_1 n_2 l_2 j_2 J T \right| {\cal G} \left|
k'_1 l_1 j_1 n_2 l_2 j_2 J T\right\rangle .
\label{hfse}
\end{equation}
The summation over the oscillator quantum numbers is restricted to the
states occupied in the IPM of $^{16}$O. This Hartree-Fock
part of the self-energy is real and does not depend on energy. One
obtains the HF single-particle wave functions by expanding them
\begin{equation}
\vert \alpha^{HF} ljm > = \sum_{i} \vert K_{i} ljm >
< K_{i} \vert \alpha^{HF}  >_{lj}
\label{hfex}
\end{equation}
in a complete and orthonormal set of regular basis functions within a
spherical box of radius $R_{box}$ which is large  compared to the
radius of the nucleus
\begin{equation}
\Phi_{iljm} ({\bf r}) = \left\langle {\bf r} \vert K_i l j m
\right\rangle = N_{il} j_l(K_ir)
{\cal Y}_{ljm} (\vartheta\varphi ) 
\label{basi}
\end{equation}
where $N_{il}$ is an appropriate normalization constant, ${\cal
Y}_{ljm}$ denotes the spherical harmonics including the spin degrees
of freedom while $j_{l}$ stands for the spherical Bessel functions
with discrete momenta $K_{i}$ determined from the boundary condition
\begin{equation}
j_l (K_i R_{\rm box}) = 0 .
\label{bess}
\end{equation}
The expansion coefficients of Eq.(\ref{hfex}) are obtained by
diagonalizing the HF Hamiltonian
\begin{equation}
\sum_{n=1}^{N}  \left\langle K_i \right| 
\frac{K_i^2}{2m}\delta_{in} +
\Sigma^{HF}_{lj} \left| K_n \right\rangle_{lj} \,
\left\langle K_n \vert
\alpha^{HF} \right\rangle_{lj} \,=  \,\epsilon^{HF}_{\alpha
lj} \,\left\langle K_i \vert \alpha^{HF}\right\rangle_{lj}. 
\label{hfha}
\end{equation}
Here and in the following the set of basis states in the box has been
truncated by assuming an appropriate $N$. From the HF wave
functions and energies one can construct the HF approximation to the
single-particle Green's function in the box, which comes out as
\begin{equation}
g_{\alpha lj}^{(HF)} (k, k'; \omega ) = \frac{<k\vert
\alpha^{HF} >_{lj}<\alpha^{HF}\vert
k'>_{lj}}{\omega -\epsilon^{HF}_{\alpha lj} \pm
i\eta} \, .
\label{hfgf}
\end{equation}
Note that by choosing especially this basis we are able to separate
contributions from different momenta to the HF single-particle state,
which is essential in order to compute at the end of our formalism the
single-particle Green's function in momentum space.

\begin{figure}[t]
\epsfysize=4.0cm
\begin{center}
\makebox[16.6cm][c]{\epsfbox{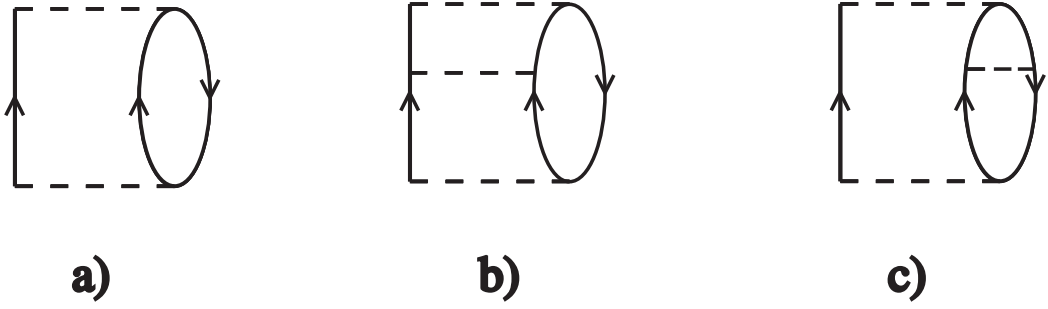}}
\end{center}
\caption{Contributions to the nucleon self-energy of second (a) and
third order in the interaction (b) and (c)}
\label{fig:diag}
\end{figure}

The next step in the evaluation of the irreducible self-energy is to
take into account terms of second order in the effective interaction
which correspond to  intermediate 2-particle 1-hole (2p1h) states as
displayed in Fig.\ref{fig:diag}a)
\begin{eqnarray}
\Sigma_{lj}^{(2p1h)}(k,k',\omega ) & = 
&  \frac{1}{2} \enspace 
\sum_{h<F} \enspace \sum_{p_{1},p_{2}>F} \enspace \frac 
{< kh \vert {\cal G}\vert p_{1} p_{2}>
<  p_{1} p_{2} \vert {\cal G}\vert k' h>}
{\omega - e(p_{1},p_{2},h) + i\eta }
\label{part}
\end{eqnarray}
and intermediate 2-hole 1-particle (2h1p) states 
\begin{eqnarray}
\Sigma_{lj}^{(2h1p)}(k,k',\omega ) & = 
&  \frac{1}{2} \enspace 
\sum_{p>F} \enspace \sum_{h_{1},h_{2}<F} \enspace \frac 
{< kp \vert {\cal G}\vert h_{1} h_{2}>
<  h_{1} h_{2} \vert {\cal G}\vert k' p>}
{\omega - e(h_{1},h_{2},p) - i\eta }\, .
\label{hole}
\end{eqnarray}
Here we have introduced the abbreviation
\begin{equation}
e(\alpha, \beta, \gamma ) = \epsilon^{HF}_\alpha + \epsilon^{HF}_\beta
- \epsilon^{HF}_\gamma
\label{hfen}
\end{equation}
where the $\epsilon^{HF}_{\alpha,\beta,\gamma}$ are the HF
single-particle energies.  Note that the summations in
Eq.( \ref{part}, \ref{hole}) on particle labels like $p_{1},p_{2}$ and
$p$ are restricted to those single-particle states within the model
space, which are above the Fermi level ($F$), whereas the labels
$h_{1},h_{2}$ and $h$ refer to hole states.

After the definition of the self-energy we can now proceed and
calculate the corresponding single-particle Green's function $g_{lj}$ by
solving a Dyson equation ( see Eq.(\ref{dyso})) with the $g^{HF}$ taken
from Eq.(\ref{hfgf}) and including the
correlation effects contained in $\Delta\Sigma_{lj}$ 
\begin{equation}
\Delta\Sigma_{lj}(k,k',\omega) = \Sigma_{lj}^{(2p1h)}(k,k',\omega)
+ \Sigma_{lj}^{(2h1p)}(k,k',\omega) \, .
\label{self}
\end{equation}

\subsection{Solution of the Dyson Equation}

The technique we use to solve the Dyson equation in order to extract
the basic ingredients of the single-particle Green's function is very
similar to the one developed in \cite{sko1,skouras,amir}. So we will
restrict ourselves in giving a short review of the basic steps towards
the determination of the single-particle Green's function for a finite
system within a model space of discrete single-particle states.

In order to obtain the information necessary for the Lehmann
representation of the Green's function, we rewrite the Dyson equation
as an eigenvalue problem \cite{sko1}
\begin{equation}
\pmatrix{H^0_{11}&\ldots &H^0_{1N} & a_{11} & \ldots & a_{1P} & A_{11} &
\ldots & A_{1Q} \cr
\vdots & & \vdots & \vdots & & \vdots &\vdots & & \vdots \cr
H^0_{N1}& \ldots &H^0_{NN} & a_{N1} & \ldots & a_{NP} & A_{N1} &
\ldots & A_{NQ} \cr
a_{11} & \ldots & a_{N1} & e_1 & & & 0 & & \cr \vdots & &\vdots & & \ddots
& & &  & \cr
a_{1P} & \ldots & a_{NP} & 0 & & e_P & & & 0 \cr
A_{11} & \ldots & A_{N1} &  & & &  E_1 & & \cr
\vdots & &\vdots& & & & & \ddots & \cr
A_{1Q} & \ldots &A_{NQ}& 0 & \ldots & 0 & & \ldots & E_Q \cr }
\pmatrix{ X_{0,k_1}^n \cr \vdots \cr X_{0,k_N}^n \cr X_1^n \cr \vdots
\cr X_P^n \cr Y_1^n \cr \vdots \cr Y_Q^n
\cr }
= \omega_n
\pmatrix{ X_{0,k_1}^n \cr \vdots \cr X_{0,k_N}^n\cr X_1^n \cr \vdots \cr
X_P^n \cr Y_1^n \cr \vdots \cr Y_Q^n
\cr } \; ,
\label{eigen}
\end{equation}
where for simplicity we have dropped the corresponding conserved
quantum numbers for parity and angular momentum ($lj$). The matrix to
be diagonalized contains the HF hamiltonian defined in
(\ref{hfha}) and the coupling to the $P$ different $2p1h$
configurations and $Q$ $2h1p$ states which can be constructed in our
model space with quantum numbers for parity and angular momentum $j$,
which are compatible to the single-particle quantum numbers $lj$ under
consideration. As long as we are still ignoring any residual
interaction between the various $2p1h$ and $2h1p$ configurations the
corresponding parts of the matrix in (\ref{eigen}) are diagonal with
elements defined by $e_i$ ($E_j$) for $2p1h$ ($2h1p$) 
\bea
e_{i} = e (p_{1}, p_{2}, h) \;& \quad & 
E_{j} = e (h_{1}, h_{2}, p) \;, 
\eea
where we have used again the abbreviation (\ref{hfen}). The matrix
elements connecting the HF part to the additional states refer to
\bea
a_{mi}  & = < k_m h \vert{\cal G} \vert p_{1} p_{2}> \nonumber \\
A_{mj}  & = < k_m p \vert{\cal G} \vert h_{1} h_{2}>
\label{mael}
\eea
Solving the eigenvalue problem (Eq.(\ref{eigen})) one
gets as a result the single-particle Green's function
in the Lehmann representation in the discrete basis of the box defined in
Eq.(\ref{basi}). The eigenvalues $\omega_n$ define the
position of the poles of the Green's function, which refer to the
various states of the system with $A$$\pm$1 nucleons and the corresponding
spectroscopic amplitudes are given by
\bea
<\Psi_0^A \vert a_{k_i} \vert \Psi_n^{A+1} > =
X_{0,k_i}^n  & \quad & for \quad \omega_n > E_F 
\nonumber \\
<\Psi_0^A \vert a_{k_i}^\dagger \vert \Psi_n^{A-1}
> =  X_{0,k_i}^n & \quad & for \quad \omega_n < E_F 
\label{spec}
\eea
which depend on whether $\omega_n$ is an energy above
or below the Fermi energy $E_F$. Note that the coefficients
$X_{0,k_i}^n$ in the above equations stand for the momentum
representation of the quasi-hole (quasi-particle) wave functions. That
means one can set $X_{0,k_i}^n=\Phi_{nlj}(k_i)$. With the help of this
nomenclature we can finally write down the spectral function in 
momentum space for a given energy $\omega_n$ and a given partial wave 
\begin{equation}
S_{lj}(k,k',\omega_n)\, =  \, X_{0,k'}^{n*} \,
X_{0,k}^n \, = \, \phi_{nlj}^*(k') \,\phi_{nlj}(k) 
\label{specfbox}
\end{equation}
in a separable representation (compare Eq.(\ref{specf})), which is
important in order to be able to use the so computed spectral function
in the description of ($e,e'p$) reactions. The corresponding
spectroscopic factor $S_n(\omega_n)$ (compare Eq.(\ref{bound})) for
the removal of a particle of a given shell $\{nlj\}$ is determined by
the norm of the quasi-hole wave function which reads according to the
chosen box basis as
\begin{equation}
S_n(\omega_n)\, = \,
\int\limits_0^{R_{box}} \!\! dr r^2 \left\vert
\Phi_{nlj}(r)\right\vert^2
\, = \,
\sum_{i} \vert X_{0,k_i}^n \vert ^2 \, ,
\end{equation}
where $\Phi_{nlj}(r)$ stands for the Fourier-Bessel transform of
$\phi_{nlj}(k)$.

In a straightforward way one can improve the approximation discussed
so far and incorporate the effects of residual interactions between
the $2p1h$ configurations as illustrated in the diagrams displaying the
self energy in Fig. \ref{fig:diag}b) and c). The same holds for the
$2h1p$ configurations.
One simply has to modify the corresponding parts of the matrix in
Eq.(\ref{eigen}) and replace
\begin{equation}
\pmatrix{e_{1} & \ldots & 0 \cr \vdots & \ddots &\cr 0 &\ldots &e_{P}\cr}
\Longrightarrow {\cal H}_{2p1h} \; ,\qquad\mbox{and}\qquad
\pmatrix{E_{1} & \ldots & 0 \cr \vdots & \ddots &\cr 0 &\ldots &E_{Q}\cr}
\Longrightarrow {\cal H}_{2h1p} \; ,
\label{restw}
\end{equation}
where ${\cal H}_{2p1h}$ and ${\cal H}_{2h1p}$ contain
the residual interactions in the $2p1h$ and $2h1p$
subspaces.  The solution of the eigenvalue problem also leads to a
normalization condition, which ensures that
\be
\sum_{n} \vert X_{0,k_i}^n \vert ^2 \, + \, \sum_{m}
\vert X_{0,k_i}^m \vert ^2  \, = \, \sum_n \vert <\Psi_n^{A+1}
\vert a^\dagger_{k_i} \vert \Psi_0^A > \vert^2 \, + \,
\sum_m \vert <\Psi_m^{A-1} \vert a_{k_i} \vert \Psi_0^A > \vert^2 
\, = \, 1 \; ,
\label{noco}
\ee
where the sum on $n$ accounts for all solutions with $\omega_n$ larger
than the Fermi energy and the sum on $m$ for all solutions with
$\omega_n$ below the Fermi energy. 
Again this implies that one ignores all effects of 
correlations, which are due to
configurations outside the model space, like {\em e.g.}\ an effective
energy-dependent hamiltonian \cite{yang,allart}. In that case
one has to renormalize the condition of
Eq.(\ref{noco}) as well.  

Note that for the solution of the eigenvalue problem one can apply the
so-called ``BAsis GEnerated by Lanczos'' (BAGEL) scheme
\cite{sko1,skouras,kuo} in order to get a very efficient
representation of the single-particle Green's function in terms of a
few ``characteristic'' poles in the Lehmann representation.

\section{Reduced Cross-sections}
\label{sec:red}

Reduced cross-sections for ($e,e'p$) reactions on $^{16}$O have
recently been measured at NIKHEFK \cite{leuschner} in the region of
missing momenta up to around 300 MeV and at MAMI in Mainz \cite{mainz}
including missing momenta up to 700 MeV.  The simplest approximation
to analyze the $(e,e'p)$ process is the Plane Wave Impulse
Approximation (PWIA), where one makes the assumption that the proton
is ejected from the nucleus without any further interaction with the
residual nucleus. In nonrelativistic PWIA the differential cross
section factorizes into two terms, the elementary electron-proton
cross section, accounting for the interaction between the incident
electron and the bound proton, and the spectral function that accounts
for the probability to find a proton with given energy and momentum in
the nucleus. Although the factorization is destroyed when one takes
into account the distortion of the electron and/or outgoing proton
waves, or a relativistic approach for the bound state, it is useful
and common practice to analyze the results in terms of a reduced cross
section defined in such a way that it would coincide with the spectral
function if factorization were fulfilled. For selected values of the
missing energy $E_m$ ({\em i.e.}\, for selected quasi-hole
excitations) the reduced cross section is given by
\begin{equation}
\rho({\bf p}_m)= \int_{\Delta E_m} dE_m \left[ \sigma^{ep}|
{\bf p}_p|E_p\right] ^{-1} \frac{d^6\sigma}{d E_p d\epsilon'_e
d\Omega_p d\Omega'_e} \; ,
\label{reduc}
\end{equation}
with ${\bf p}_m$ the experimental missing momentum ($=-P_{A-1}$),
$E_p,|{\bf p}_p|,\Omega_p$ ($\epsilon'_e, \Omega'_e$) the outgoing
proton (electron) kinematical variables, and experimentally, the
integral is performed over the interval $\Delta E_m$ that contains the
peak of the transition under study. The term $\sigma^{ep}$ represents
the elementary electron-proton cross section. The data of $\rho({\bf
p}_m)$ are obtained dividing the experimental cross section by
$\sigma^{ep}_{cc1}$, as given by Eq.(17) of Ref.~\cite{deF83}. We
therefore use the same expression for $\sigma^{ep}$ in our theoretical
calculations. In PWIA, $\rho({\bf p}_m)$ represents the momentum
distribution of the selected quasi-hole state $\alpha$. 

In this section we briefly summarize the formalism used to
describe the ($e,e'p$) reaction. More details can be found in Refs.
\cite{Jin,Udietal,Udi93}. We base our calculations on the impulse
approximation (virtual photon absorbed by the detected nucleon),
which is known \cite{Fru84} to be a reliable approximation at
quasi-elastic kinematics.
The basic equations that determine
the reduced cross section are given explicitly in
Refs. \cite{Jin,Udietal}, in terms of the electron and nuclear
currents. The calculations have been performed with the code developed
by one of us \cite{Udithesis}.

We work in the laboratory frame in which the target nucleus is at
rest and use the notation and conventions of Ref. \cite{BD64}. We
denote by $k_e^\mu=(\epsilon_e,{\bf k}_e)$ the four--momentum of the
incoming electron and by ${k'}_e^\mu=(\epsilon'_e,{\bf k'}_e)$ the
four--momentum of the outgoing one. The four--momentum of the
exchanged photon is $q^\mu=k_e^\mu-{k'}_e^\mu=(\omega,{\bf q})$.
$P_A^\mu=(M_A,{\bf 0})$ and $P_{A-1}^\mu =(E_{A-1}, {\bf P}_{A-1})$
denote the four--momenta of the target and residual nucleus, while
$p_p^\mu=(E_p,{\bf p}_p)$ is the four--momentum of the ejected
proton.

Using plane waves for the electrons and considering knock--out from a
given $\{nlj\}$ quasi-hole state, we write the amplitude for the
$(e,e'p)$ process in DWIA as \cite{Udietal,Udi93,Fru84}:
\begin{equation} 
W_{if}=\frac{m_e}{\sqrt{\epsilon_e \epsilon'_e}} \bar{u}({\bf
k'}_e,\sigma'_e)\gamma_\mu u({\bf k}_e,\sigma_e) \frac{(-1)}{q_\mu^2} J_N^\mu
(\omega,{\bf q})\; , \label{wif} 
\end{equation}
where $u({\bf k},\sigma)$ represent four--component relativistic free
electron spinors \cite{BD64}, and $J_N^\mu (\omega,{\bf q})$ is the
nuclear current \begin{equation} J_N^\mu(\omega,{\bf
q})=\overline{\sum_I} \sum_F\delta(E_F-E_I-\omega)
\int\!\!\!  d{\bf r} e^{i{\bf q}\cdot {\bf r}} <\Psi^F_{A} | \hat{J}^\mu _N|
\Psi^I_A> \; , \label{jmucurrent} \end{equation} where the matrix element of
the nuclear charge-current density operator is taken between the
initial $|\Psi^I_A>$ and the final $|\Psi_A^F>$ nuclear states. The
initial state will be given by the many-particle wave function of A
bound particles in the ground state for the target nucleus
$|\Psi_0^A>$. For the final state, the experimental conditions dictate
a state that behaves asymptotically as a knocked out nucleon with
momentum $p_p$ and a residual nucleus in a well-defined state
$|\Psi_n^{A-1}(E)>$ with energy $E$ and quantum numbers $n$. It is
possible \cite{polls,boffireport} to describe the matrix
element in Eq. (\ref{jmucurrent}) in terms of the simple one-body
current operator sandwiched between the hole spectral functions
\begin{equation} 
[S_n]^{\frac{1}{2}}
\phi_{E_n}({\bf p})= 
<\Psi_n^{A-1}(E)|a({\bf p})|\Psi^A_0>
\label{bound}
\end{equation}
and
\begin{equation}
\chi^{(-)}_{p_p E_n}=<\Psi_n^{A-1}(E)|a({\bf p})|\Psi^F_A>
\label{continuum}
\end{equation}
describing the overlap between the residual state $|\Psi_n^{A-1}(E)>$
and the hole produced in $|\Psi^A_0>$, and $|\Psi^F_A>$ respectively,
by removing a particle with momentum ${\bf p}$. The way to compute
these overlaps was described in the previous section.  Hence, we take
$\phi_{En}$ normalized to 1, and $S_n(E)$ is the spectroscopic factor
associated with the removal process.

One may perform a similar calculation for $\chi^{(-)}_{p_p E_n}$, the
final (continuum) wave funciton of the ejected proton. Usually,
however, this particle wave function is derived from a
phenomenological local optical potential. Also in this work we
compute the the wave function for the outgoing proton $\chi^{(-)}_{p_p
E_n}$ as a scattering solution obtained from an optical potential in a
relativistic framework fitted to elastic proton scattering data on
$^{16}$O \cite{hama}. No Perey factors are included, as the dependence
on the energy of these potentials is very soft at the energies of
interest.

To avoid expansions in p/M and the reductions of the current operator
typical of the nonrelativistic approaches, the fully relativistic
kinematics and structure of the operator is used. In order to do that,
we build a 4-component spinor out of the nonrelativistic bispinor
$\phi_{E_n}$. Also, we use relativistic optical potentials
\cite{hama}, which according to previous results \cite{Udietal,highp},
seem to be more adequate for $(e,e'p)$ calculations. However, our
calculation is essentially nonrelativistic and, in consistency with
the derivation of the spectral functions, no contributions from the
negative energy sector are allowed in neither the bound or scattered
proton wave function. The results would be equivalent to the ones
obtained with the nonrelativistic formalism and equivalent optical
potentials, only that all orders of p/M are included in our
calculation. We obtain similar results with a nonrelativistic
prescription for the scattered states (nonrelativistic optical
potentials and Perey factors), only the reduced cross-sections are
slightly increased.

For the nucleon current operator we take the free nucleon expression
\begin{equation}
\hat{J}_{N}^\mu =F_1\gamma^\mu+i \frac{\bar{\kappa}F_2}{2M}
\sigma^{\mu\nu}q_\nu \; ,
\label{cc2}
\end{equation}
where $F_1$ and $F_2$ are the nucleon form factors related in the
usual way \cite{BD64} to the electric and magnetic Sachs form factors
of the dipole form. As discussed in Refs. \cite{Udietal,Chin}, DWIA
results depend on the choice of the nucleon current operator. Here we
have chosen the operator that is closer to the one used in the
nonrelativistic calculations of the {\sc dweepy} code, and most often
employed.

The Coulomb distortion of the electron wave function is considered
with an effective momentum approach. The value of the shift in the
energy of the electron is obtained by comparing with the exact
calculation (DWBA)\cite{Udi93} and it is set in this case to about 4
MeV. For a light nuclei as $^{16}$O, this procedure is already
sufficient in order to include this effect accurately in the
cross-sections \cite{JinO16}.

\section{Results and discussion}
\label{sec:res}

\begin{figure}
\begin{center}
\epsfysize=7.0cm
\makebox[16.6cm][c]{\epsfbox{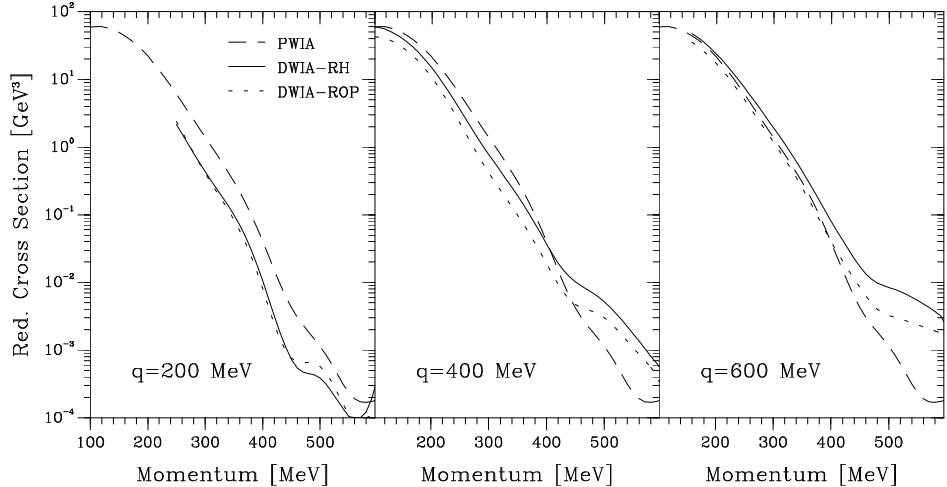}}
\end{center}
\caption{Reduced cross section for the $^{16}$O($e,e'p$)$^{15}$N
reaction leading to the ground state of $^{15}$N in perpendicular
kinematics. The results for PWIA are compared with those from DWIA
calculated with a relativistic Hartree potential (RH) and a relativistic
optical potential (ROP).}
\label{fig:md}
\end{figure}

In the first part of this discussion we would like to
emphasize the effects of the final state interaction for the outgoing
proton.  For this purpose Fig.~\ref{fig:md} displays the reduced cross
section, calculated in perpendicular kinematics for the
$^{16}$O($e,e'p$)$^{15}$N reaction leading to the ground state
($1/2^-$) of $^{15}$N, as a function of missing momentum. Results are
shown for three different momentum transfers $q$ of the virtual
photon. As a reference all three parts of this figure include the
results of the PWIA. The PWIA yields a reduced cross section which is
identical to the spectral function and independent on the momentum of
the absorbed photon. Two different models have been employed to
describe the distortion of the wave function of the outgoing
proton. The first model is based on a microscopic
Dirac-Brueckner-Hartree-Fock calculation for the nucleus $^{16}$O
\cite{fritz}. This DWIA-RH (Relativistic Hartree) approach only
yields a real part for the nucleon nucleus potential. The second
approach (DWIA-ROP) also accounts for the absorption of the outgoing
proton by means of inelastic scattering. Also the DWIA-ROP approach
assumes a relativistic form of the otical potential. The scalar and
vector terms for the real and imaginary part are obtained in a
phenomenological way by a global fit of elastic proton-nucleus
scattering data
\cite{hama}.

With the inclusion of the final state interaction, the cross section
cannot be factorized any more into the free electron-nucleon cross
section times a spectral function. Therefore the reduced cross
sections displayed in Fig.~2 depend on the momentum $q$ of the
absorbed photon. The difference between the two-models of the final
state interaction can mainly be attributed to the effects of the
imaginary component in DWIA-ROP. The absorptive effects contained in
this imaginary part leads to a reduction of the cross section (as
compared to the DWIA-RH results) which seems to be rather insensitive
with respect to the missing momentum $p_m$. Therefore in the following
we will restrict the presentation of calculated cross sections to the
DWIA-ROP approach.

The real part of the optical potential tends to smoothen
the reduced cross section by shifting strength from small momenta to high
missing momenta. Independent on the detailed form of the optical potential one
also observes an increase of the reduced cross section with increasing photon
momentum $q$.

After exploring these features of the final state interaction, we now want to
compare our results for the reduced cross sections with experimental data.
In particular we would like to study how the effects of long-range correlations
may effect the shape of the spectral function.
 The long-range correlations are
described in terms of the 2p1h and 2h1p configurations within a model
space including single-particle states up to the sdg shell and taking
into account effects of the residual interaction between these
configurations. For a more detailed discussion on the importance of
contributions to the self-energy represented by diagrams of third and
higher order in the residual interaction as displayed by Figs.
\ref{fig:diag}b and \ref{fig:diag}c and their effects on the momentum
distribution and other observables we refer to Ref. \cite{amir}. 

\begin{figure}
\begin{center}
\epsfysize=7.0cm
\makebox[16.6cm][c]{\epsfbox{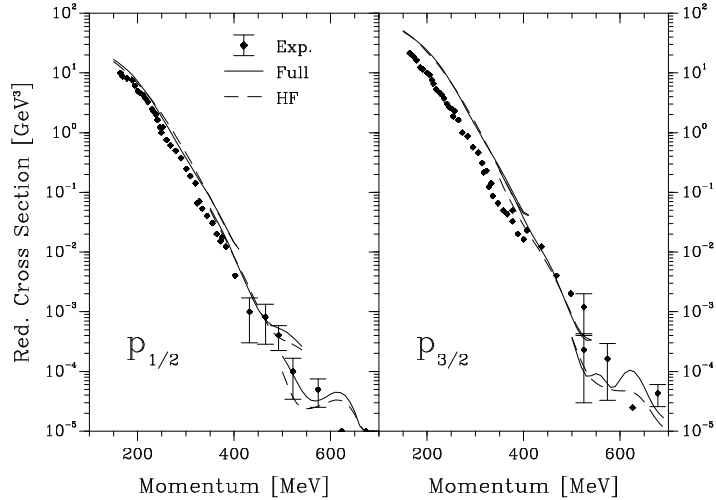}}
\end{center}
\caption{Reduced cross section for the $^{16}$O($e,e'p$)$^{15}$N
reaction leading to the ground state ($1/2^-$) and first excited state
($3/2^-$) of $^{15}$N in the kinematical conditions considered in the
experiment at MAMI (Mainz) \protect\cite{mainz}. Results for the
mean-field description (HF) and the fully correlated spectral function
(Full) are presented. The spectroscopic factors determined by a fit to
low-p data from the NIKHEF experiment \protect\cite{leuschner} are
$S_{0p_{1/2}}$=0.60 (0.83) and $S_{0p_{3/2}}$=0.45 (0.85), where the
values obtained by the theoretical approach are enclosed in parenthesis.}
\label{fig:hm}
\end{figure}

To observe the influence of long-range correlations on the momentum
distribution, we present in Fig.~\ref{fig:hm} the reduced cross section
for the $^{16}$O($e,e'p$)$^{15}$N reaction leading to the ground state
($1/2^-$) and first excited state ($3/2^-$) of $^{15}$N. These have been 
calculated both with the fully correlated spectral function (Full) and also 
with our mean-field
(HF) description (see Eq.(\ref{hfgf})). The results are compared with
the experimental data taken from the experiment at MAMI (Mainz)
\cite{mainz}. The underlying formalism to compute the reduced cross
section is performed as described previously, where the
phenomenological relativistic optical potential \cite{hama} was
used. The relevant piece of the nuclear structure calculation for the
proton knock-out reaction is the quasi-hole part of the fully
correlated spectral function for the $p_{1/2}$ and $p_{3/2}$ partial
wave presented in the previous chapter (see Eq.(\ref{specfbox}) and
Eq.(\ref{restw})). As  we are mainly interested in the comparison
of the data at high missing momenta with the predictions at low momenta, 
the corresponding spectroscopic factors $S_n(E)$ 0.60 ($0p_{1/2}$) and 0.45 
($0p_{3/2}$) were determined by a fit of the calculated cross section to the
experimental results at small missing momenta deduced 
from the NIKHEF data \cite{leuschner}. These adjusted spectroscopic factors 
are significantly smaller than the theoretical values of 0.83 and 0.85 for the
$p_{1/2}$ and $p_{3/2}$ states, respectively. For this comparison, however, one
must keep in mind that these calculated spectroscopic factors only account for
the effects of long-range correlations. Short-range correlations should lead to
additional reduction by another factor of 0.8 and subnucleonic degrees of
freedom may be responsible for the remaining discrepancy.

As one can clearly see in Fig.~\ref{fig:hm}, the inclusion of
long-range correlations does not lead to a significant enhancement of
the reduced cross section at high missing momenta as compared to the
predictions of the mean field or Hartree-Fock approach. This
conclusion is not altered even in a complete description of the
($e,e'p$) experiment including FSI and other effects. This statement
confirms the result already obtained by an investigation of long-range
correlations ({\em i.e.}\ correlations induced by excitation modes at
low energies), on the energy and momentum distribution for the nucleus
$^{16}$O \cite{amir}. It should be emphasized, however, that a rather
reasonable description of the reduced cross section is obtained over a
large range of missing momenta with the adjustment of only one
parameter, the spectroscopic factor, which has been adjusted to
describe the NIKHEF data at small missing momenta.

\begin{figure}
\begin{center}
\epsfysize=7.0cm
\makebox[16.6cm][c]{\epsfbox{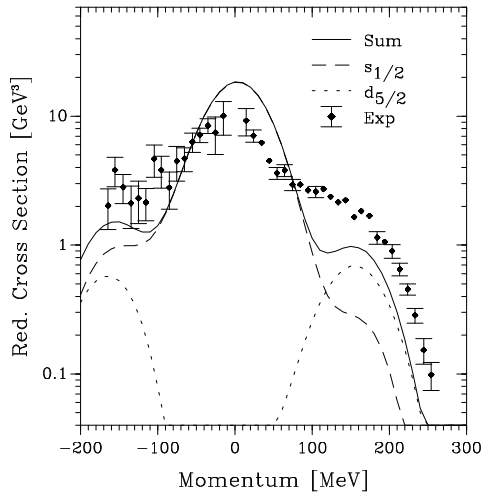}}
\end{center}
\caption{Reduced cross section for the $^{16}$O($e,e'p$)$^{15}$N
reaction leading to the first excited state with positive parity
($1/2^+-5/2^+$-dublett) of $^{15}$N in parallel kinematics. The
experimental data are taken from NIKHEF experiment
\protect\cite{leuschner}. The corresponding spectroscopic factors
$S_{s_{1/2}}$=0.055 and $S_{d_{5/2}}$=0.035 are taken from the
theoretical approach.}
\label{fig:leu}
\end{figure}

Now one could argue that in the ($3/2^-$) and ($1/2^-$) case the
spectral function has a dominant quasi-hole part originating from the
corresponding hole ({\em i.e.}\ bound) state. Therefore any effects of
correlations would be overwhelmed by this dominating quasi-hole
component, which is present already in the Hartree-Fock
approximation. This argument is supported by observing that in
Fig.~\ref{fig:hm} the reduced cross section computed with the fully
correlated spectral function does not differ substantially from the
mean field description. In this sense it would be more interesting to
investigate transitions to final states which do not have such a
dominant quasi-hole part and would be impossible to describe
within the mean field
approximation. Best examples at low missing energies in $^{16}$O is
the case of the $d_{5/2}$ and $s_{1/2}$ hole states.  As we are able
to compute the fully correlated spectral functions also for these
partial waves it would be worth comparing these results with
corresponding experimental data. Unfortunately, until now we are not
aware of data from the MAMI (Mainz) experiment \cite{mainz} for high
missing momenta of these partial waves.  So we had to take the
experimental data from the NIKHEF experiment in parallel kinematics
\cite{leuschner} which cover only the momentum region up to 280
MeV/c. Moreover, they could not resolve the contributions from the
($5/2^+$) and ($1/2^+$) states. In Fig.~\ref{fig:leu} we display the
results obtained for the reduced cross section taking the fully
correlated spectral functions and assuming an incoherent sum of the
$1/2^+$ and $5/2^+$ contributions for the $^{16}$O($e,e'p$)$^{15}$N
reaction leading to the first excited state with positive parity
($1/2^+-5/2^+$-dublett) of $^{15}$N. The experimental data points are
taken from the NIKHEF experiment \cite{leuschner} mentioned above.

Contrary to the previous case, the curves in Fig.~\ref{fig:leu} are
scaled with the spectroscopic factors given by the theoretical
approach, 0.055 ($s_{1/2}$) and 0.035 ($d_{5/2}$) respectively.  The
reasonable agreement with experiment shown in Fig.~\ref{fig:leu},
means that the underlying microscopic calculation of the spectral
function could be used to explore the high momentum region.  Note that
one could get better agreement with the experimental results by using
the spectroscopic factors of 0.0357 ($s_{1/2}$ and 0.1140 ($d_{5/2}$
which have been obtained by fitting the experimental data \cite{leuschner}.
In view of our aim to test our theoretical approach
this has not been done here. 

\section{CONCLUSIONS}
\label{sec:conc}

The reduced cross section for the ($e,e'p$) process has been studied
in DWIA for the example of the nucleus $^{16}$O using a spectral
function containing long-range correlations. The fully correlated
spectral function in the various partial waves is derived from the
single-particle Green's function, which is obtained from the solution
of a Dyson equation in basis of plane-wave states. The self-energy for the
nucleons entering this Dyson equation contains the
Brueckner-Hartree-Fock term plus the coupling to 2p1h and 2h1p
configurations inside a model space with single-particle states up to
the $sdg$ shell. The long-range correlations are described in terms of
these 2p1h and 2h1p configurations including the residual interaction
among them.

Effects of final state interactions for the outgoing protons are included by
means of relativistic optical model interactions. The effects of the final 
state  interaction are non-negligible and  spoil the factorization of the total
cross section into the spectral function times the free electron-nucleon cross
section. The effects of the final state interaction turn out to be rather
insensitive to the specific choice of the optical model.

Including these effects of the final state interaction the calculated cross
section for $(e,e'p)$ reactions leading to the $0p_{1/2}$ and $0p_{3/2}$
quasi-hole states in $^{16}$O agree rather well with the experimental data
\cite{mainz} over a large range of missing momenta. It turn out, however, that
these data are rather insensitive to correlation effects.  Spectral functions
derived from the Hartree-Fock approach yield an agreement of similar quality.

In order to investigate the effects of correlations, one should study $(e,e'p)$
reactions to final states, which are impossible within the mean field or
Hartree-Fock approach. Examples for such states at low missing energies are the
$5/2^+$ and $1/2^+$ states in $^{15}$N. The cross section derived from the
correlated single-particle Greens function is in good agreement with the
experimental data.

\section{ACKNOWLEDGEMENTS}

This project has been  supported from the EC-program `Human
Capital and Mobility' under Contract N. CHRX-CT 93-0323. It has also received
support by the ``Sonderforschungsbereich SFB 382''.

\end{document}